\documentclass[%
 reprint,
 amsmath,amssymb,
 aps,
]{revtex4-2}

\usepackage{graphicx}
\usepackage{dcolumn}
\usepackage{bm}
\usepackage{xspace}
\usepackage{xcolor}
\usepackage{float}
\usepackage[mathlines]{lineno}

\newcommand{\figref}[1]{Fig.~\ref{#1}}
\newcommand{\Figref}[1]{Figure~\ref{#1}}
\newcommand{\tabref}[1]{Tab.~\ref{#1}}

\newcommand{\secref}[1]{Sec.~\ref{#1}}

\newcommand{\NeCOtwoNtwo}{Ne-CO$_2$-N$_2$ (90-10-5)\xspace}

\begin{document}
\preprint{APS/123-QED}

\title{A new method to search for highly ionizing exotic particles, monopoles and beyond, using time projection chamber}

\author{Mesut Arslandok$^{1}$, Helen Caines$^{1}$ and Marian Ivanov$^{2}$}
 \affiliation{%
 $^{1}$Wright Lab, Physics Department, Yale University, New Haven, CT 06520, USA \\
 $^{2}$GSI Helmholtzzentrum für Schwerionenforschung, 64291 Darmstadt, Germany
}%

\date{\today}

\begin{abstract}

Measuring the energy loss and mass of highly ionizing particles predicted by theories from beyond the Standard Model pose considerable challenges to conventional detection techniques. Such particles are predicted to experience energy loss to matter they pass through that exceeds the dynamic range specified for most readout chips, leading to saturation of the detectors' electronics. Consequently, achieving precise energy loss and mass measurements becomes unattainable. We present a new approach to detect such highly ionizing particles using time projection chambers that overcomes this limitation and provide a case study for triggering on magnetic monopoles.

\end{abstract}

\maketitle


\section{\label{sec:intro}Introduction}

The discovery of highly ionizing particles (HIP) predicted from theories beyond the Standard Model, such as monopoles, gluinos, Q-balls, strangelets, heavy leptons, etc., would address a number of important questions in modern physics, including potentially the origin and composition of dark matter in the universe and the unification of the fundamental forces~\cite{Nath:2010zj,ParticleDataGroup:2006fqo,Perl:2001xi,Fairbairn:2006gg}. While most predictions suggest that such particles are far too massive to be produced in any foreseeable accelerator~\cite{Mavromatos:2020gwk,tHooft:1974kcl}, other models suggest a few of these particles, such as magnetic monopoles, could occur in the mass range accessible via the Large Hadron Collider (LHC) at CERN~\cite{Fairbairn:2006gg}. Various technologies and techniques are already being used at the LHC to search for such particles either through tracking detectors by the ATLAS and CMS collaborations~\cite{ATLAS:2015tyu,ATLAS:2012bda,ATLAS:2023esy,HassanElSawy:2744867}, or the use of passive detectors, such as the studies underway within the MoEDAL collaboration~\cite{MoEDAL:2014ttp,MoEDAL:2016jlb,MoEDAL:2016lxh,MoEDAL:2017vhz,MoEDAL:2019ort,MoEDAL:2021vix,Mitsou:2022cuw}. However, detecting HIPs, especially in tracking detectors, poses significant challenges. For example, the passage of a HIP through a tracking detector is expected to lead to energy deposits that typically cause saturation of readout electronics, making a positive identification considerably more difficult~\cite{CMSTracker:2005iuk,ALEPH:1997siq}. 

We propose a novel approach for detecting HIPs that utilizes the continuous readout currently made possible with gas electron multiplier (GEM)-based time projection chambers (TPCs)~\cite{Sauli:1997qp,Sauli:2014cyf} to overcome this saturation effect. Such TPCs are also robust against discharges, even in scenarios with extremely high energy depositions. In this innovative method the TPC acts as a passive detector by incorporating an algorithm into the readout electronics that uses the so-called common-mode (CM) signal ~\cite{ALICETPC:2020ann,ALICETPC:2023ojd} -- a negatively polarized signal below the baseline -- as a hardware trigger, followed by offline track reconstruction for HIP identification. The intrinsic design of a GEM-TPC setup offers other distinct advantages, such as a low material budget, compared to solid-state tracking detectors, and an extensive 3-D tracking volume.


This report is organized as follows. In~\secref{sec:ALICETPC} we give a brief introduction to TPCs and provide more details on the common-mode effect. In~\secref{sec:monopoles}, as an example of our proposed HIP identification technique, we conduct an in-depth case study on potential energy loss measurements of magnetic monopoles using the GEM based TPC that is currently in operation at the ALICE experiment at the LHC~\cite{ALICE:2008ngc,ALICE:2014sbx}. Finally we conclude in~\secref{sec:Conclusions} with a discussion on how the proposed approach can be employed for more general HIP identification.

\section{\label{sec:ALICETPC} GEM-based TPCs, the common-mode effect, and ALICE's TPC}

TPCs~\cite{Nygren:1974wta,Nygren:1984ljp,Charpak:1973mug,Sauli:2014cyf} are typically built as large gas filled cylindrical or box-shaped chambers. Examples of TPCs in current use in nuclear physics heavy-ion experiments include in NA61~\cite{NA61:2014lfx} at the SPS, STAR~\cite{STAR:1997sav,STAR:1992kiw} and sPHENIX~\cite{Klest:2022yrp} at RHIC, and ALICE~\cite{Alme:2010ke,ALICETPC:2020ann} at the LHC. Equipped with an array of sensitive detectors (e.g. MWPCs and GEMs), TPCs capture the ionization signals generated by charged particles passing through the gas. As particles move through the gas, they ionize the atoms or molecules along their path, generating ionization electrons and positively charged ions. 
The electrons are collected at a readout plane via an electric field that is applied across the gas volume. An additional large electric field at the MWPC or GEM readout plane creates an electron avalanche leading to a significant increase in the number of electron-ion pairs and thus the signal amplitude. The arrival points in the readout plane provide the $x$ and $y$ coordinates, while the drift time is used to determine the $z$-coordinate of the initial ionization point. In this way TPCs enable reconstruction of the trajectories of charged particles in three dimensions. The typical resolution along the time axis ($z$) is coarser than solid-state trackers, being typically on the order of several hundred microns compared to ten microns. However, TPCs benefit from their low material budget and abundance of ionization points, enabling segmentation of the x-y plane into many more samples as compared to solid trackers. This capability allows for the detection of long tracks and effective use in large volumes, making TPCs well suited to high track density environments. In addition, the ability to control the amplification enables energy loss measurements of HIPs.

The primary challenge for TPCs is posed by the ions generated during both primary ionization and the subsequent gas amplification. These ions induce a so-called space charge effect, which distorts the electric field within the gas volume and influences the trajectory of subsequent primary ionization electrons. The amount of ions produced during gas amplification far exceeds that of primary ionization, so TPCs are designed to prevent as many ions as possible from entering the drift volume. Multi-wire proportional chambers (MWPCs)~\cite{Sauli:2014cyf} employ gating wires for this purpose. These wires are positioned as the outermost wire layers and operate in two modes: open and closed. In open gate mode, all wires of the gating grid are kept at a common potential so that the grid is transparent to charge transport between amplification and drift regions. In closed mode, on the other hand, the ions are blocked with alternating voltages, resulting in an ion leakage into the drift region of less than a factor of 10$^{-4}$~\cite{Alme:2010ke}. The gating grid is closed by default and only opened in case of an event trigger. This opening and closing of the gating wires imposes limitations on TPC readout speeds. In contrast, GEMs~\cite{Sauli:1997qp,Sauli:2014cyf} inherently block ions from reentering the drift region without the need for a gating grid, albeit with a higher rate of ion leakage. In the ALICE TPC, for example, an ion leakage of about 0.7\% was achieved by optimizing the voltage settings and hole sizes of the GEM foils~\cite{ALICETPC:2020ann}. The continuous readout operation mode of GEM-based TPCs enables the measurement of very long signal tails, caused for example by the significant energy depositions expected from HIPs - a capability we will further discuss below in the context of the ALICE TPC.

The ALICE TPC, which is operated in a 0.5~T solenoidal magnetic field parallel to its axis, is the main tracking and particle identification (PID) detector in the central barrel of the ALICE experiment~\cite{ALICE:2000jwd,Alme:2010ke,ALICETPC:2020ann}. The TPC consists of a large cylindrical vessel, with a total active volume of 88 $m^3$, filled with a \NeCOtwoNtwo gas mixture (i.e. 90 parts of Ne, 10 parts of CO$_2$, and 5 parts of N$_2$)~\cite{ALICETPC:2020ann}. It is divided equally into two drift regions by a negatively charged central membrane at its axial center. The TPC sectors, each covering 20$^{\circ}$ in azimuth, are located at the ends of the drift volume. They are radially segmented into inner and outer readout chambers (18 IROC and 18 OROC on each side respectively). A uniform electric field along the z-axis is generated by the field cage. 
The anode plane of the readout chambers comprises a total of 524,160 pads, which are arranged in 152 rows in the radial direction. The number of pad rows determines the largest possible number of hit points -- so-called clusters, concentrated charge deposits that are detected within a search window extending over 3 bins in the pad-row direction and 3 bins in the time direction -- along the trajectory of a given particle. Each pad is connected to a front-end electronics channel with 160 front-end channels combined in each front-end card. More details about the TPC and its performance can be found in Refs.~\cite{Alme:2010ke,ALICETPC:2020ann}.

The ALICE TPC was upgraded from MWPC-based readout to GEM-based readout during the second Long Shutdown (December 2018--March 2022) of the LHC. This upgrade was necessary to cope with the predicted minimum bias Pb--Pb collision rate of 50~kHz during the Run3 and Run4 (2022--2030) data acquisition periods~\cite{ALICETPC:2020ann,ALICE:2012dtf}. To achieve the required gain while effectively suppressing the back-flow of ions produced during the amplification stage, a quadruple GEM configuration was chosen. A TPC sector is divided into four GEM stacks: one for IROC and three for OROC (see~\figref{fig:monopole_track}) in the radial direction. The main components of the mechanical structure of the readout chambers consist of a trapezoidal aluminium frame (Al-body), a pad plane separated from the Al-body by a fiberglass plate (strongback), and a stack of four GEM foils, more details of the structure of an IROC are shown in Fig.~6 of~\cite{ALICETPC:2020ann}.

\begin{figure}[h]
\includegraphics[width=8.5cm]{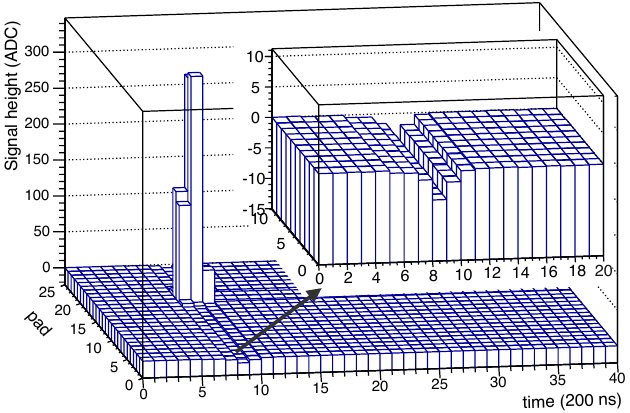}
\caption{\label{fig:CM_signal} The main figure shows the ADC signal after pedestal subtraction and before zero suppression for a hit in an ALICE GEM-TPC sector, the inset shows the common-mode effect measured at the same time along the other pads. Figure taken from \cite{ALICETPC:2020ann}.}
\end{figure}
\begin{figure}[h]
\includegraphics[width=1\linewidth]{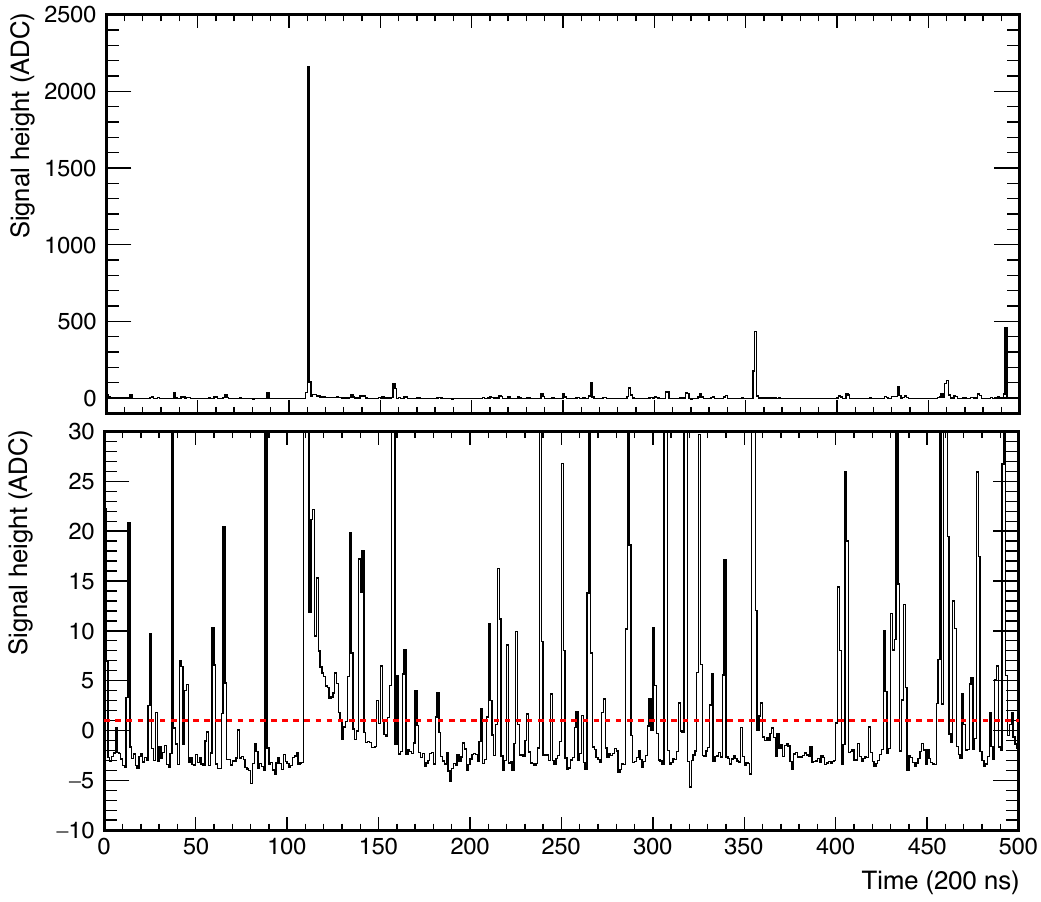}
\caption{\label{fig:run3_baseline} Simulation of a pad signal for an event with $\approx$30\% occupancy, representing the highest expected track density in the ALICE TPC in Run3. Top: full signals are shown. Bottom: a zoom in on the $y$-axis to reveal the baseline fluctuations, the red-dashed line shows the zero-suppression threshold. Figure taken from~\cite{ALICETPC:2023ojd}.}
\end{figure}

Within the GEM readout system, the pad plane and the GEM foils possess an inherent capacitance that leads to the so-called common-mode effect~\cite{ALICETPC:2023ojd}. It is created by their common high-voltage supply through a resistor network. It is worth noting that a similar common-mode effect is observed in the MWPC-based ALICE TPC~\cite{Arslandok:2308311}. When a signal is detected on a single pad, a capacitive signal with opposite polarity is induced across all pads facing the corresponding stack. Therefore, the amplitude of the common-mode signal in a given pad is suppressed by a factor of $N_{\rm pads}$ with respect to the original signal, where $N_{\rm pads}$ is the number of pads belonging to the same stack. \Figref{fig:CM_signal} shows a typical semi-Gaussian TPC signal together with the simultaneous common-mode signal seen as an undershoot in the neighboring pads. As a result, the common-mode effect leads to an average baseline drop and an effective noise contribution, as illustrated via simulation in~\figref{fig:run3_baseline}. At the highest expected occupancy during Run3 and Run4, the average baseline shift of the ALICE GEM-based readout is about 3~ADC counts, compared to an average signal peak height of about 80~ADC counts. 

In the following, we discuss the expected response of the ALICE TPC in the presence of an energy deposit equivalent to that of a monopole. We show that the amplitude of the baseline drop is expected to be much larger providing inspiration for use as a monopole or any other HIP trigger.

\section{\label{sec:monopoles}HIP identification: A case study on Monopoles} 

\begin{figure*}[ht!]
   \centering
   \includegraphics[width=0.85\linewidth]{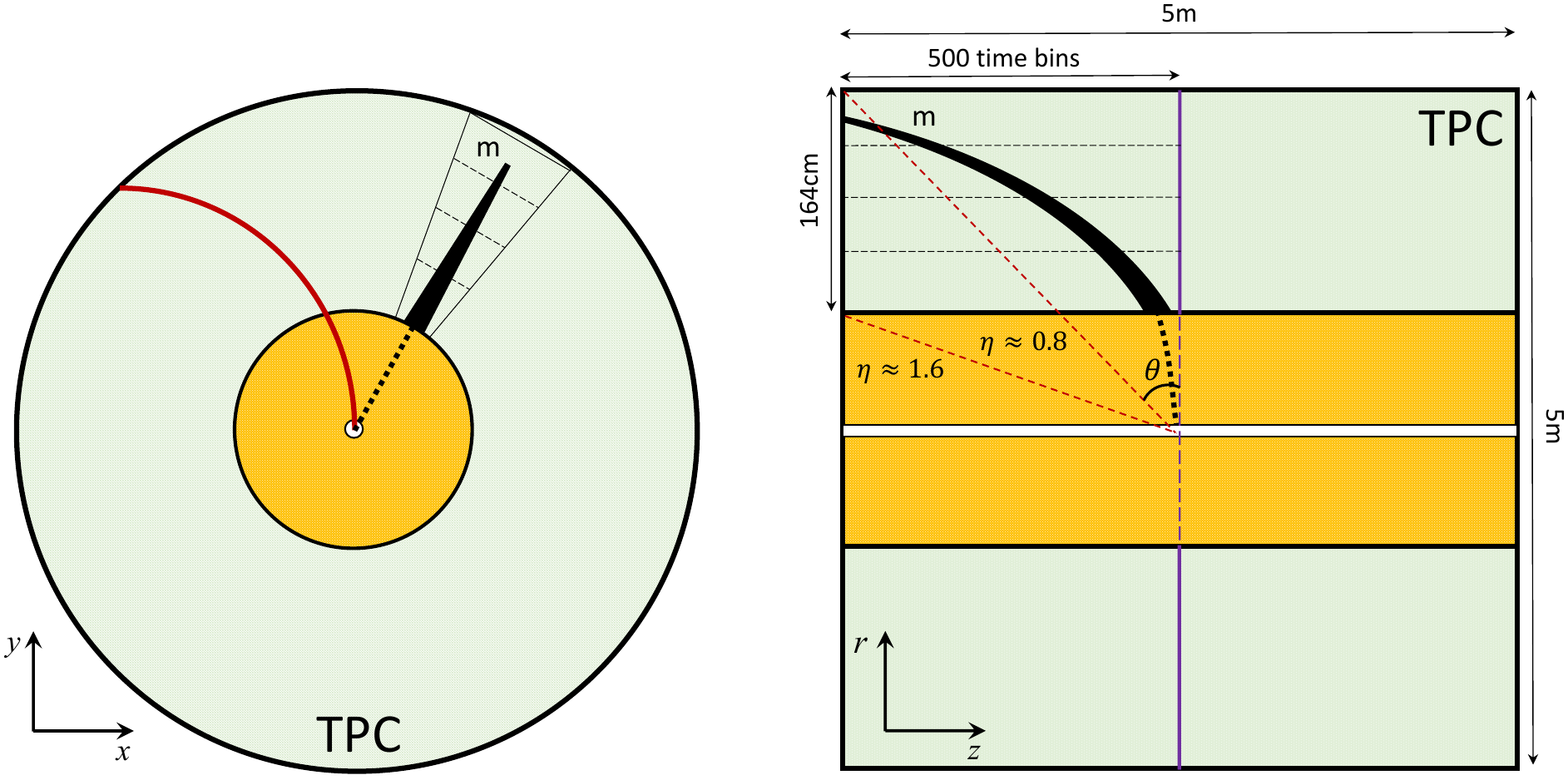}
   \caption{\label{fig:monopole_track} (Color online) Left: the $x$-$y$ ($r$-$\phi$) projection of the ALICE TPC. The black line denotes a monopole (m) trajectory on a particular TPC sector, its thickness in the green volume labeled ``TPC", indicates the expected energy loss. The red curve represents the helix-like trajectory of a charged particle. Right: the $r$-$z$ view highlights the parabolic trajectory of the monopole. The black dashed lines represent the segmentation of the four GEM stacks, while the red dashed line at $\eta\approx0.8$ marks the point where the track length on the $x$-$y$ plane begins to decrease, continuing until $\eta\approx1.6$, which represents the border of the geometric acceptance of the TPC. These drawings are simple sketches and are not drawn to scale. }
\end{figure*}
\begin{figure*}[ht!]
   \centering
   \includegraphics[width=0.85\linewidth]{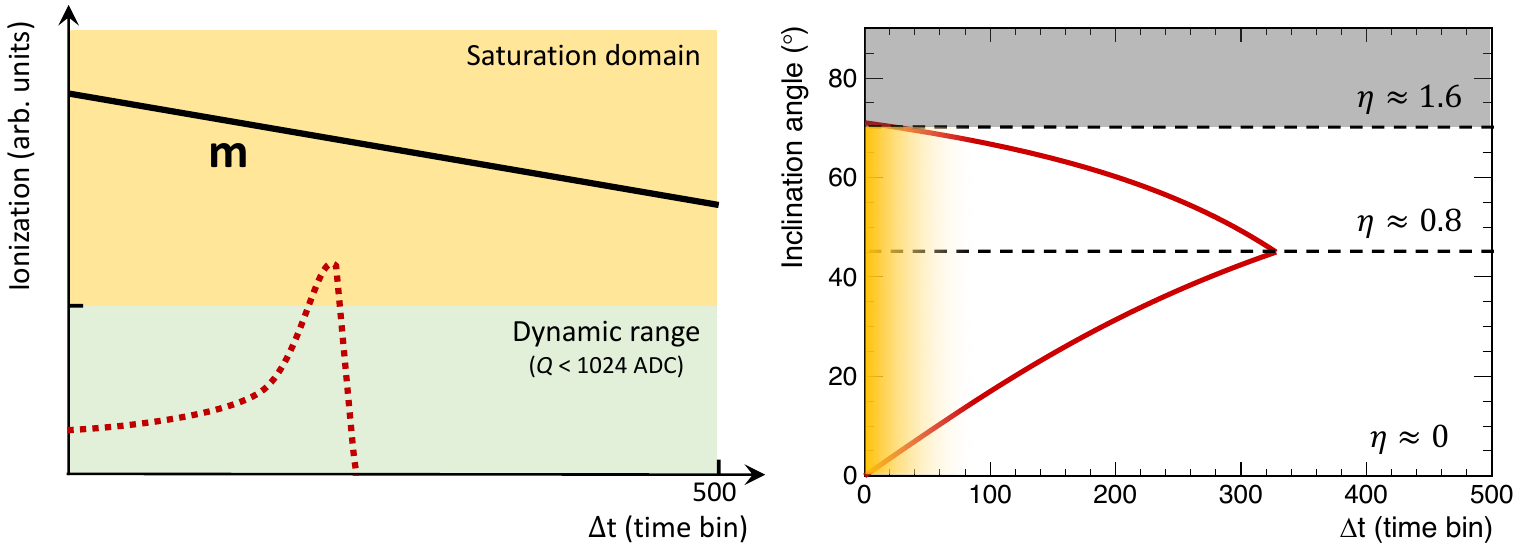}
   \caption{\label{fig:monopole_deltaT} (Color online) Left: expectations of the energy loss over time within the TPC volume for a monopole compared to a typical high-z particle, shown as solid black and dashed red curves, respectively. In the gas, a Bragg peak is observed when the high-z particle stops, while for the monopole the ionization becomes less dense as it loses energy. The green shaded area marks the ionization levels within the dynamic range of the readout electronics where the detected charge per channel remains below 1024 ADC. The yellow area highlights where saturation of the channels occurs. Right: the projection of a HIP track in the saturation domain onto the time ($z$) axis shown as a function of inclination angle ($\theta$), assuming no bending of the track due to the magnetic field (indeed, monopoles with a large mass ($>\sim$100 GeV) and low charge would leave a straight track consistent with a very high-momentum electrically charged particle~\cite{Fairbairn:2006gg}). Tracks with inclination angles smaller than 45 degrees (i.e., $\eta < 0.8$) exhibit larger $\Delta t$ values when they bend under the influence of the magnetic field, as depicted in Figure \ref{fig:monopole_track}. The gray shaded area shows the angles outside the geometric acceptance of the TPC, while the orange shaded area with a color gradient indicates where the background for the HIPs, mainly spallation products, is anticipated and should be investigated experimentally.}
\end{figure*}
When a relativistic monopole~\cite{Mavromatos:2020gwk} possesses a single Dirac charge and is in motion, its energy loss due to ionization is equivalent to that of a relativistic particle carrying an electric charge with $|z|\approx68.5$, where $z$ stands for the number of elementary charges in units of `$e$'. Consequently, the ionization caused by a relativistic monopole in matter is expected to exceed the ionization caused by a minimum-ionizing particle (MIP) by more than a factor of 4700. In the presence of a magnetic field, it also exhibits two unique properties. Firstly, it follows a parabolic trajectory~\cite{Fairbairn:2006gg,TASSO:1988txx,Bauer:2005xc}, diverging from the anticipated helix-like trajectory of an electrically charged particle moving in a magnetic field. In addition, the energy loss of the monopole decreases as it slows down, whereas an electrically charged particle exhibits the opposite trend, resulting in a Bragg peak when it stops. Both of these characteristics are demonstrated in \figref{fig:monopole_track}, which depicts the trajectories of a monopole and a charged particle traversing the gas medium of the ALICE TPC in both the $x$-$y$ and $r$-$z$ planes. 

\begin{table*}[ht]
	\centering
	\begin{tabular}{|c|l|c|}
		\hline
		\multicolumn{1}{|c|}{\textbf{Variable}} & \multicolumn{1}{c|}{\textbf{Definition}} & \multicolumn{1}{c|}{\textbf{Value}} \\
		\hline
		$N_{\mathrm{pads}}$    & Number of pads in an IROC                 & 5280 \\
		$N_{\mathrm{rows}}$    & Number of rows in an IROC                 & 63 \\
		$k_{\mathrm{CF}}$      & Common-mode fraction factor for an IROC   & 0.42 \\
		$Q_{\mathrm{max,MIP}}$ & Mean maximum charge for a MIP signal      & 20~ADC \\
	    $k_{\mathrm{gain}}$    & Number of electrons produced per primary electron during gas amplification & 2000 \\	
        $N_{\mathrm{el,MIP}}$  & Number of primary electrons per MIP       & 40 \\ 
        ${L_\mathrm{pitch}}$   & Distance between GEM holes                & 140~$\mu$m \\
        ${L_\mathrm{pad}}$     & Pad length in IROC                        & 1~cm \\
		\hline
	\end{tabular}
	\caption{Some ALICE TPC specific variables~\cite{ALICETPC:2023ojd,ALICETPC:2020ann}.}
	\label{tab:tpcvariables}
\end{table*}
\begin{figure*}[ht]
   \centering
   \includegraphics[width=0.8\linewidth]{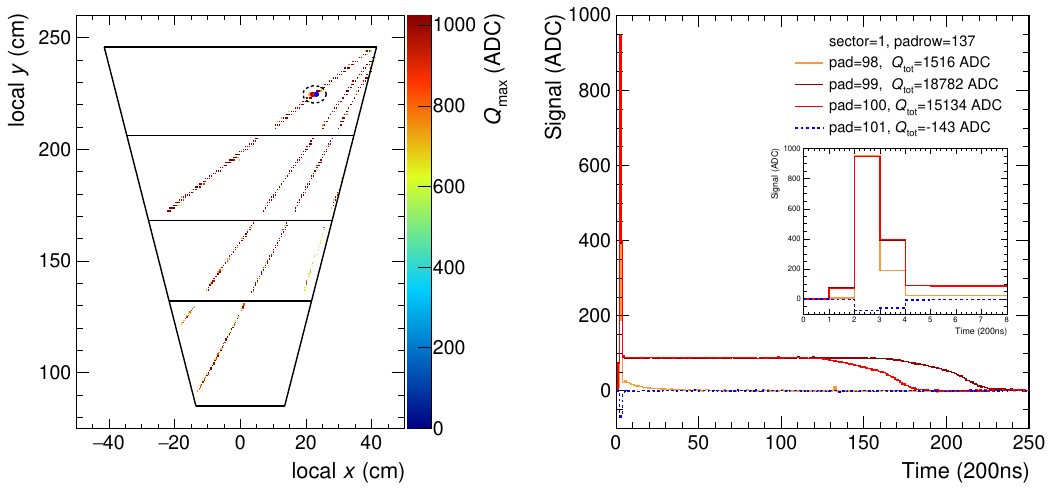}
   \caption{\label{fig:saturated_signal}Left: three highly ionizing laser tracks in a given TPC readout chamber projected onto the local $x$-$y$ plane. Radial segmentation of the GEM stacks is also shown with horizontal solid lines ordered from bottom to top: IROC (stack~0), OROC~1 (stack~1), OROC~2 (stack~2), and OROC~3 (stack~3). Right: saturated signals of a cluster, shown by the dashed circle in the left panel. Note the stable operation of the detector in such high energy depositions. Figure taken from~\cite{ALICETPC:2023ojd}. }
\end{figure*}
Figure \ref{fig:monopole_deltaT} compares the energy loss and track length in timebins in the TPC gas between a monopole and a typical high-z particle, in this case spallation products, which constitute the primary background for the HIP search. The duration for which a monopole track remains in the saturation domain of the readout chip is comparatively longer. Additionally, not only are the characteristics of the energy loss different, but also the magnitude is expected to be relatively larger in the case of monopoles. Here, it is important to note that monopoles with masses of about 100~GeV are anticipated to be generated through the Schwinger pair production mechanism during ultraperipheral heavy-ion collisions~\cite{Gould:2019myj,Gould:2021bre,MoEDAL:2021vix,Mitsou:2022cuw}. These collisions are understood to produce the strongest known magnetic fields in the universe, with minimal background interference. 

Ultimately, the discovery of a monopole using this new technique demands the identification of a single, very highly ionizing parabolic track with decreasing energy loss along its path. The remaining key question - assuming that the monopole survives the material before it reaches the TPC gas - is whether the TPC and its readout can handle such high-energy depositions. We address this question in more detail below.

The signal response of the detector using laser calibration data has been recently published by the ALICE TPC collaboration in Ref.~\cite{ALICETPC:2023ojd}. In Fig.~\ref{fig:saturated_signal} three highly ionizing laser tracks and the corresponding signals of a cluster are depicted. In the following, we examine how these signals can be used to test a monopole scenario. The right panel of Fig.~\ref{fig:saturated_signal} shows that signals that are excessively large saturate the dynamic range of the SAMPA chip~\cite{ALICETPC:2020ann}, such that the sum of the pedestal value and the measured charge ($Q$) cannot exceed 1024~ADC. This poses a challenge for accurately measuring the signal amplitude. However, using the measured common-mode signal (shown by the blue dashed curve with the total and maximum charge of $Q_{\text{tot,CM}}=-143$~ADC and $Q_{\text{max,CM}}=-80$~ADC, respectively), which is proportional to the total charge recorded in the stack, one can estimate the mean energy loss for a given cluster (measured as MIP equivalent) with the following equation:
\begin{align}
Q_{\mathrm{max,Laser}} = \frac{ [(Q_{\mathrm{max,CM}} \times N_{\mathrm{pads}})/N_{\mathrm{rows}}] \times (1/k_{\mathrm{CF}}) }{ Q_{\mathrm{max,MIP}} },
\label{eq:QmaxLaser}
\end{align}
where:
\begin{itemize}
     \item{$Q_{\mathrm{max,CM}} \times N_{\mathrm{pads}}$ is the total maximum charge in IROC,}
     \item{$N_{\mathrm{rows}}$ is the normalization factor to obtain the energy loss for a given hit point,}
     \item{$k_{\mathrm{CF}}$ is the common-mode fraction factor, which is defined as the ratio of the total negative charge to the total positive charge in a given stack (table~2 of Ref.~\cite{ALICETPC:2023ojd}),}
     \item{$Q_{\mathrm{max,MIP}}$ is used as the normalization factor to get a MIP equivalent value.}
\end{itemize}

By way of demonstration we substitute into this equation the values for the ALICE IROC, see~\tabref{tab:tpcvariables}, and determine a MIP equivalent value of 798 for the laser track in Fig.~\ref{fig:saturated_signal}. This is approximately six times lower than the MIP equivalent of 4700 determined above for a monopole. It is worth noting that this value applies when the monopole is oriented perpendicular to the beam axis. However, monopoles in a magnetic field are expected to bend along a parabolic trajectory~\cite{Fairbairn:2006gg,TASSO:1988txx,Bauer:2005xc}, and thus the deposited charge per monopole cluster is distributed across multiple time bins. More importantly, as illustrated in~\figref{fig:monopole_deltaT}, the number of time bins, i.e. $\Delta t$, for a given monopole track with significant common-mode signals will be comparatively large compared to spallation products. 

Another important constraint that must be considered is the probability of discharge under conditions of high ionization. This was investigated in Ref~\cite{Gasik:2017uia}, where the reaction of a single GEM setup with gas mixtures based on Argon and Neon on alpha particles was studied. Discharge occurs when the total accumulated charge within a single GEM hole exceeds the critical charge limit ($Q_{\mathrm{crit}}$). For a Ne-CO$_2$ (90-10) gas mixture, which can be taken as a reasonable proxy to the original gas mixture of \NeCOtwoNtwo, $Q_{\mathrm{crit}}$ was determined to be $(7.3 \pm 0.9) \times 10^{6}$ electrons per GEM hole.

To estimate the number of electrons per hole in the IROC geometry~\cite{ALICETPC:2020ann} for a monopole, one can proceed as follows
\begin{align}
N_{\mathrm{el, Mon}} = \frac{[N_{\mathrm{el, MIP}} \times Q_{\mathrm{HIP}} \times k_{\mathrm{gain}}]}{ \frac{2\sigma_{\mathrm{D}}}{L_{\mathrm{pitch}}} \times \frac{L_{\mathrm{pad}}}{L_{\mathrm{pitch}}}},
\label{eq:NElMono}
\end{align}
where:
\begin{itemize}
     \item{$[N_{\mathrm{el,MIP}} \times Q_{\mathrm{HIP}} \times k_{\mathrm{gain}}]$ is the total number of electrons per pad after gas amplification, the MIP-equivalent energy loss of a monopole is $Q_{\mathrm{HIP}}=4700$,}
     \item{$2\sigma_{\mathrm{D}}/L_{\mathrm{pitch}}$ accounts for the diffusion of electrons over 5~mm distance ($\sigma_{D}\approx2.5$~mm for a flat diffusion assumption in 1~m electron drift),}
     \item{$L_{\mathrm{pad}}/L_{\mathrm{pitch}}$ accounts for the number of holes per pad length.}
\end{itemize}
Using this equation, the calculation yields an electron count per hole of $1.2 \times 10^{5}$ for an ALICE GEM stack, notably one order of magnitude lower than the critical charge, $Q_{\mathrm{crit}}$. In addition with the four-GEM configuration of ALICE the total gain is distributed among the foils building the stack, resulting in increased stability against discharges. Therefore, this value serves as a conservative lower limit. However, large local charge densities have the potential to trigger sparks that lead to high-voltage trips. Consequently, data acquisition for the chamber is temporarily halted until the high voltage is restored. This interruption begins within a millisecond, while the complete recovery process of the high-voltage power supply takes a few minutes. On the other hand, the signal induction on the pads for the whole track segment occurs on a time scale of less than a microsecond. As a result, even if the signal tail above the baseline is cut off, the common-mode signal below the baseline remains detectable, ensuring that the particle trajectory is measurable.

So far, we have demonstrated the effective performance of GEM-based TPCs in extreme energy loss scenarios. In the following, we outline our systematic proposal for reconstructing the HIP tracks:
\begin{itemize}
     \item{\textbf{Hardware trigger in the FPGA:} Two specific pieces of information are required to trigger the monopole candidates: first, for a given pad a high common-mode signal that exceeds a trigger threshold $Q_{\rm tr}$ must be detected. $Q_{\rm tr}$ is set significantly higher than the average baseline shift. Second, a count of subsequent time bins exceeding $Q_{\rm tr}$ (denoted as $\Delta t$ in \figref{fig:monopole_deltaT}) is needed to determine the track length along the z-axis. When $\Delta t$ is sufficiently large, the pad and time bin of the initiating trigger is flagged to record the signal without baseline subtraction along with the rest of the event data for further analysis. As noted above $\Delta t$ of a monopole is expected to be significantly larger than that of spallation products and Bragg peaks, allowing effective rejection of background signals. This type of trigger offers two distinct advantages: first, operating at the hardware level enables analysis across any collision system as well as cosmic events. Second, it is sensitive to short track lengths along both the radial and time directions that can not be identified by typical track reconstruction algorithms. Such algorithms struggle to track particles crossing only a few pad rows and usually assume helical trajectories. It is important to emphasize that the parabolic trajectories of low-momentum monopoles are expected to cross only a few pad rows due to their bending along the z-axis, as shown in~\figref{fig:monopole_track}. Note also that the common-mode signal within a given time bin cannot exceed the pedestal value, which ranges from 40-120 ADC in the ALICE TPC, with an average value of $\sim$80 ADC (see Fig.~44 of ref.~\cite{ALICETPC:2020ann}), since charge measurements in ADC counts are strictly positive (see Fig.~50 of Ref.~\cite{ALICETPC:2020ann}). This limitation might complicate the accurate measurement of extreme energy loss cases where the common-mode signal would exceed the pedestal value. Nevertheless, the particle trajectory and a lower limit of the energy loss via integration of the part of the common-mode signal that is recorded can still be determined.}
     \item{\textbf{Raw data recording:} accurate tracking and energy loss measurement requires raw data without subtracting the baseline. The recorded time window must be long enough to cover the entire track candidate ensuring that both the common-mode signals and the signals above the baseline, including long saturation tails, are captured (see blue dashed and solid red curves in~\figref{fig:saturated_signal}, respectively). Note that in the case of MWPCs, i.e. triggered readout, the extended saturation tail would be cut off.}
     \item{\textbf{Cluster finding:} cluster finding algorithm for signals above the baseline should include the entire signal tail for precise energy loss measurement. In the case of a hardware trip, the common-mode signal can still be used as a trigger due to its fast response time, as can the measurement of energy loss, even if the signal tail is distorted.}
     \item{\textbf{Tracking:} a special tracking algorithm is required to handle both electrically (helix topology) and magnetically (parabolic topology) charged tracks. This algorithm should be applied to the clusters above the baseline to determine the trajectory, while the measurements of the total charge of these clusters are used together with the common-mode signal to accurately measure the energy loss. This step is particularly important to reject background using the track topology and length information.}
\end{itemize}
In an optimal detector configuration, the measurement of common-mode signals and thus the detection of HIPs can be improved by fine-tuning experimentally adjustable parameters. These parameters include $k_{\mathrm{gain}}$ (the multiplication factor of the primary ionization, adjustable through voltage settings on the GEM foils), $N_{\mathrm{pads}}$ (proportional to the capacitance, i.e. GEM stack area facing the pad plane), the number of layers, the distance to the collision point, the magnetic field strength, the material budget before the TPC, the dynamic range of the readout chips and the pedestal values of the readout channels. 

Accordingly, the recipe for the ideal TPC for HIP detection is summarized as follows. The $k_{\mathrm{gain}}$ parameter should be set low, e.g. to 100 instead of 2000, to increase the sensitivity to HIPs while suppressing signals from all other standard-model particle species including electrons, muons, pions, kaons, protons, etc.. By significantly reducing the recorded track density this adjustment would also help mitigate the space charge generated in the amplification region. Such conditions would be suitable for any TPC designed for HIP detection across various interaction rate scenarios. 

By optimizing the high voltage segmentation of the GEM foils, the capacitance between the pad plane and the GEM stack can be adjusted to ensure that the common-mode signal does not exceed the pedestal. Additionally, increasing the number of GEM layers would enhance stability against discharge~\cite{Gasik:2017uia}. Minimizing the material budget, which involves avoiding additional detectors before the TPC and reducing the material budget for the beam pipe, will maximize the probability that the monopole survives and enters the active volume of the TPC. This prevents the HIPs from stopping before reaching the TPC and also reduces the distance between the TPC and the beam pipe, thereby enhancing the kinematic acceptance. For example, one could consider utilizing the geometry of the yellow area instead of the green in~\figref{fig:monopole_track}, which enables a more compact TPC design with a geometric coverage close to $4\pi$ due to the minimal distance to the collision point. Moreover, increasing the magnetic field to 2 T instead of 0.5 T would increase track bending and thus increase the track length along the z-axis, i.e. the number of common-mode triggers. Expanding the dynamic range of the readout chip beyond 1024~ADC enables the measurement of larger energy depositions. Finally, setting a high pedestal value, e.g. 1000~ADC, will increase the dynamic range of the HIP's common-mode signal thereby facilitating precise energy loss measurements.

\section{\label{sec:Conclusions}Conclusions}

This article presents a new approach utilizing time projection chambers to detect highly ionizing particles such as monopoles. As an illustration, we examined the response of the ALICE TPC to energy depositions corresponding to those of a Dirac monopole, equivalent to 4700~MIPs. Our findings indicate that the measured signals, corresponding to an energy loss of about 798~MIPs, are promising to achieve this goal. Additionally, we have estimated the number of electrons per hole under the conditions of 4700~MIPs and found it to be about an order of magnitude below the critical charge density, where discharges occur. 
Moreover, we investigated the possibility that in cases where detector stability is compromised, reconstruction of the common-mode signal remains a viable option for tracking and measuring the energy loss of the highly-ionizing particles. 

After validating the stable operation of the GEM-based ALICE TPC under significant energy depositions, we have introduced an innovative method in which the TPC functions as a passive detector. This method utilizes the negatively polarized common-mode signal as a hardware trigger in the readout electronics, enabling the recording of digitized raw data without loss of information, which is crucial for the precise energy loss and mass measurements. We then elaborated on a TPC-based detector design that, in combination with this technique, can serve as an optimal detector offering complete $4\pi$ coverage for the study of highly ionizing particles beyond the Standard Model. Such a detector is well-suited for both high and low interaction rate scenarios and offers significant advantages in the discovery of HIPs, in particular monopoles in background-free ultraperipheral heavy ion collisions.

\begin{acknowledgments}
We would like to thank Chilo Garabatos, Robert Helmut M\"{u}nzer and Christian Lippmann for their valuable discussions regarding the ALICE TPC and its stability, and to Torsten Alt, Peter Hristov, David Rohr and Ruben Shahoyan for their insights on the data acquisition system of the ALICE detector. This work is supported in part by the US DOE under award number DE-SC004168, the Bundesministerium f\"{u}r Bildung und Forschung (BMBF) and GSI Helmholtzzentrum f\"{u}r Schwerionenforschung GmbH, Germany;
\end{acknowledgments}

\nocite{*}

\bibliography{apssamp}

\end{document}